\documentclass[a4paper, 11pt]{article}
\usepackage{a4wide}
\usepackage{amsmath}
\usepackage{commath}
\usepackage{latexsym}
\usepackage{graphicx}
\usepackage{color}
\usepackage[pdfstartview=FitH]{hyperref}
\usepackage{verbatim}
\usepackage{authblk}
\usepackage[T1]{fontenc}
\usepackage{cite}

\title{Empirical Survival Jensen-Shannon Divergence as a Goodness-of-Fit Measure for Maximum Likelihood Estimation and Curve Fitting}

\author{Mark Levene mark@dcs.bbk.ac.uk}
\affil{Department of Computer Science and Information Systems, \authorcr Birkbeck, University of London, London WC1E 7HX, U.K.}
\author{Aleksejus Kononovicius aleksejus.kononovicius@tfai.vu.lt}
\affil{Institute of Theoretical Physics and Astronomy, \authorcr Vilnius University, Vilnius, Lithuania}

\date{}

\begin{document}

\maketitle

\begin{abstract}

The coefficient of determination, known as $R^2$, is commonly used as a goodness-of-fit criterion for fitting linear models. $R^2$ is somewhat controversial when fitting nonlinear models, although it may be generalised on a case-by-case basis to deal with specific models such as the logistic model. Assume we are fitting a parametric distribution to a data set using, say, the maximum likelihood estimation method. A general approach to measure the goodness-of-fit of the fitted parameters, which is advocated herein, is to use a nonparametric measure for comparison between the empirical distribution, comprising the raw data, and the fitted model. In particular, for this purpose we put forward the {\em Survival Jensen-Shannon divergence} ($SJS$) and its empirical counterpart (${\cal E}SJS$) as a metric which is bounded, and is a natural generalisation of the Jensen-Shannon divergence. We demonstrate, via a straightforward procedure making use of the ${\cal E}SJS$, that it can be used as part of maximum likelihood estimation or curve fitting as a measure of goodness-of-fit, including the construction of a confidence interval for the fitted parametric distribution. Furthermore, we show the validity of the proposed method with simulated data, and three empirical data sets.

\end{abstract}


\setlength{\parskip}{4pt}

\section{Introduction}

We assume a general scenario, where we have some data from which we derive an empirical distribution that is
fitted with maximum likelihood \cite{MYUN03} or curve fitting \cite{FOX16} to some, possibly parametric distribution \cite{KRIS15}.

The {\em coefficient of determination}, $R^2$ \cite{MOTU95}, is a well-known measure of goodness-of-fit for linear regression models.
Despite its wide use, in its original form, it is not fully adequate for nonlinear models, \cite{ANDE94}, where the author recommends to define $R^2$ as a comparison of a given model to the null model, claiming that this view allows for the generalisation of $R^2$.
Further, in \cite{SPIE10} the inappropriateness of $R^2$ for nonlinear models is clearly demonstrated via a series of Monte Carlo simulations.
In \cite{CAME97}, a novel $R^2$ measure based on the Kullback-Leibler divergence \cite{COVE06} was proposed as a measure of goodness-of-fit for regression models in the exponential family.
In addition, in \cite{OREL08} problems with using $R^2$ for assessing goodness-of-fit in linear mixed models with random effects were highlighted, and in \cite{NAKA13} an improved extension was proposed in the context of both linear and generalised linear mixed models. Despite numerous proposals to address the issues with $R^2$ for nonlinear models, many of them are ad-hoc, and, as noted in \cite{SAPR14}, should be applied with caution. In summary, there seems to be a lack of general purpose goodness-of-fit measures that could be applied to nonlinear models, which is the main issue we attempt to redress with the ${\cal E}SJS$.

Alternative nonparametric methods have also been proposed. In particular, the Akaike information criterion ($AIC$) and its counterpart the Bayesian information criterion ($BIC$) \cite{BURN02,VRIE12}, are widely used estimators for model selection.
Both AIC and BIC are asymptotically valid maximum likelihood estimators, with penalty terms to discourage overfitting.
We stress that goodness-of-fit measures how well a single model fits the observed data, while model selection compares the predictive accuracy of two models relative to each other \cite{BURN02,PITT02}.
The likelihood ratio test is also an established method for model selection between a null model and an alternative maximum likelihood model \cite{VUON89,LEWI11}.
Despite the popularity of maximum likelihood methods, there is some controversy in their application as goodness-of-fit tests \cite{HEIN03}.

The {\em Jensen-Shannon divergence} ($JSD$) \cite{LIN91,ENDR03}  is a symmetric form of the nonparametric Kullback-Leibeler divergence \cite{COVE06}, providing a measure of distance between two probability distributions. It has been employed in a wide range of applications such as
detecting edges in digital images \cite{GOME00}, measuring the similarity of texts \cite{MEHR15}, training adversarial neural networks \cite{GOOD14},
comparison of genomes in bioinformatics \cite{SIMS09}, distinguishing between quantum states in physics \cite{MAJT05}
and as a measure of distance between distributions in a social setting \cite{FENN16b}.

Here we generalise the $JSD$ to the {\em survival Jensen-Shannon divergence} ($SJS$) and it empirical counterpart (${\cal E}SJS$) by employing the survival functions of the constituent probability density functions. We apply the ${\cal E}SJS$ as an alternative measure of goodness-of-fit of a parametric distribution, acting as the model, to an empirical distribution, which  comprises the raw data. The ${\cal E}SJS$ provides a direct measure of goodness-of-fit without the need of the maximum value of the likelihood function, as used in the AIC and BIC, or any linearity assumptions of the model being fitted, as often made when using $R^2$.

The rest of the paper is organised as follows.
In Section~\ref{sec:jsd}, we introduce the {\em empirical survival Jensen-Shannon Divergence} (${\cal E}SJS$) and some of its characteristics.
In Section~\ref{sec:gof}, we define the ${\cal E}SJS$ as a measure of goodness-of-fit within the context of distribution fitting and define the notion of the ${\cal E}SJS$ factor. In Section~\ref{sec:exp}, we describe some experiments we carried out, with simulated data in Subsection~\ref{subsec:simulated} and empirical data in Subsection~\ref{subsec:empirical}, to test the viability of using the ${\cal E}SJS$ as a measure of goodness-of-fit.
Finally, in Section~\ref{sec:conc}, we give our concluding remarks.

\section{Survival Jensen-Shannon Divergence}
\label{sec:jsd}

First, in Subsection~\ref{subsec:entropy} we define the survival function with respect to the {Lebesgue\!--\!Stieltjes} measure, and then the empirical survival entropy, which generalises the entropy to continuous distributions. Building on these concepts, we then define, in Subsection~\ref{subsec:sjs} the survival Jensen-Shannon divergence, which generalises the standard Jensen-Shannon divergence ($JSD$), and derive a formula for computing its empirical counterpart.

\subsection{Survival Entropy}
\label{subsec:entropy}

We define the {\em survival function} $S = S_X$, for real-valued random variable $X$ as
\begin{displaymath}
S_X(x) = P \left( X > x \right),
\end{displaymath}
which represents the probability that $X$ takes a value greater than $x$.
It can be shown that there exists a unique probability measure $P_S$ for $S$, with respect to the $\sigma$-field of Borel sets on the real line, defined as
\begin{equation}\label{eq:leb-st}
P_S((a,b]) = S(a) - S(b).
\end{equation}

The measure in (\ref{eq:leb-st}) is known as the {\em Lebesgue\!--\!Stieltjes} measure;
the detailed theory can be found in, for example, \cite{CAPI04} and \cite{SHOR17} and an informative resource describing the various types of integral and induced measures can be found in \cite{BURK07}.

Let $X_1, X_2, \ldots, X_n$ be $n$ independent and identically distributed random variables forming a random sample drawn from a population having survival function $S$. The {\em empirical survival function}, denoted by $\hat{S}_n$, is given by
\begin{equation}\label{eq:emp}
\hat{S}_n(x) = \frac{1}{n} \sum_{i=1}^n I_{\{X_i > x\}},
\end{equation}
where $I$ is the indicator function. It is well known, according to the Glivenko-Cantelli theorem, that $\hat{S}_n$ converges to $S$ as $n$ tends to infinity; for the details see, for example \cite{SHOR17}.

The {\em empirical survival entropy}, (called the empirical cumulative entropy in \cite{RAO04}), is given by
\begin{equation}\label{eq:se1}
{\cal E}(\hat{S}_n) = - \int_{0}^{\infty} \hat{S}_n(x) \log \hat{S}_n(x) \ {\rm d}x,
\end{equation}
using the convention that $0 \log 0 = 0$.

Following \cite{DICR09} let $0 = X_{(0)} \le X_{(1)} \le X_{(2)} \le \ldots \le X_{(n)}$ be the order statistics of the random sample, implying that
\begin{equation}\label{eq:se2}
{\cal E}(\hat{S}_n) = - \sum_{i=1}^{n-1} \int_{X_{(i)}}^{X_{(i+1)}} \hat{S}_n(x) \log \hat{S}_n(x) \ {\rm d}x.
\end{equation}

Moreover,
\begin{displaymath}
\hat{S}_n = 1- \frac{i}{n}, \ \ {\rm when} \ \ X_{(i)} \le x < X_{(i+1)}, \ \ {\rm for} \ \ i = 1,2,\ldots,n-1,
\end{displaymath}
and thus on using (\ref{eq:se1}) it follows that
\begin{equation}\label{eq:se3}
{\cal E}(\hat{S}_n) = - \sum_{i=1}^{n-1} U_{i+1} \left( 1 - \frac{i}{n} \right) \log \left( 1 - \frac{i}{n} \right),
\end{equation}
where $U_i = X_{(i)} - X_{(i-1)}$ is the sample spacing of the random sample drawn from a population having survival function $S$; see also \cite{YARI12}.

Again, by the Glivenko-Cantelli theorem, it can be shown that ${\cal E}(\hat{S}_n)$ would converge
to ${\cal E}(S)$ as $n$ tends to infinity, where ${\cal E}(S)$ is obtained by replacing $\hat{S}_n$ in (\ref{eq:se1}) by $S$; see \cite{RAO04,DICR09}.
As a result, here we will concentrate on the methodology and empirical analysis of large data sets, and leave the study of its application to small data sets as an open problem.

\subsection{Survival Jensen-Shannon Divergence}
\label{subsec:sjs}

Let $P$ and $Q$ be survival functions of the density function $p$ and $q$, respectively, and let $M$ be the survival function of the mixture $\frac{1}{2} (p+q)$. We define the {\em Survival Jensen-Shannon divergence} (SJS), as
\begin{equation}\label{eq:js1}
SJS(P,Q) = \frac{1}{2} \int_0^\infty \ \left( P(x) \log \frac{P(x)}{M(x)} + Q(x) \log \frac{Q(x)}{M(x)} \right) {\rm d} x.
\end{equation}

It is easy to see that
\begin{equation}\label{eq:js2}
SJS(P,Q) = \int_0^\infty \ \left( \frac{1}{2} P(x) \log P(x) + \frac{1}{2} Q(x) \log Q(x) -  M(x) \log M(x) \right) {\rm d} x.
\end{equation}

The {\em empirical SJS} is thus given by
\begin{equation}\label{eq:ejs1}
{\cal E}SJS(P,Q) = \frac{1}{2} \int_0^\infty \ \left( \hat{P}_n(x) \log \frac{\hat{P}_n(x)}{\hat{M}_n(x)} + \hat{Q}_n(x) \log \frac{\hat{Q}_n(x)}{\hat{M}_n(x)} \right) {\rm d} x,
\end{equation}
which by (\ref{eq:js2}) can be stated as
\begin{equation}\label{eq:ejs3}
{\cal E}SJS(P,Q) = - \int_0^\infty \ \left( \frac{1}{2} {\cal E}(\hat{P}_n) + \frac{1}{2}  {\cal E}(\hat{Q}_n) - {\cal E}(\hat{M}_n)  \right) {\rm d} x.
\end{equation}

Finally, on using (\ref{eq:se3}) we have
\begin{equation}\label{eq:ejs4}
{\cal E}SJS(P,Q) = \sum_{i=1}^{n-1} \left( \frac{1}{2} U[P]_{i+1} + \frac{1}{2} U[Q]_{i+1} - U[M]_{i+1} \right) \left( 1 - \frac{i}{n} \right) \log \left( 1 - \frac{i}{n} \right),
\end{equation}
where $U[P]_{i}$, $U[Q]_{i}$ and $U[M]_{i}$ are, respectively, the sample spacings drawn from populations having the survival functions $P$, $Q$ and $M$.

An important fact to note is that the square root of the ${\cal E}SJS$ is a metric \cite{NGUY15}, generalising the result for the standard $JSD$ \cite{ENDR03}.
It is also bounded but with a different normalisation constant than that of the $JSD$. Moreover, we observe that it is often advantageous, as we do here, to use the survival function (or equivalently the cumulative distribution) instead of the probability density function as it may be easier to interpret and manipulate, and it also acts to smooth the data.
We further note that we are dealing with continuous distributions and therefore it makes more sense to use the ${\cal E}SJS$, based on the survival function, rather than on the empirical counterpart of the $JSD$ based on the density function, which is only defined at discrete points.
In this context we also note that our use of the empirical survival distribution is compatible with the empirical distribution function statistics as in \cite{STEP86}, which measure the absolute difference between the empirical distribution and an assumed distribution from the data it is sampled from.

\section{Empirical Jensen-Shannon Divergence as a Goodness-of-Fit Measure}
\label{sec:gof}

Making use of the ${\cal E}SJS$ as a measure of goodness-of-fit is quite straightforward.
Assume that $P(D, \phi)$ is a sample from a parametric distribution $D$, with parameters $\phi$,
and that $D$ is fitted with maximum likelihood \cite{MYUN03} or curve fitting \cite{FREU06} to an empirical distribution, $\hat{P}$.

The goodness-of-fit of the distribution $D$, with parameters $\phi$, to the empirical distribution $\hat{P}$ is now defined as
\begin{equation}\label{eq:gof}
{\cal E}SJS(P(D,\phi), \hat{P}),
\end{equation}
where $P(D, \phi)$ is a sample from the distribution $D$ with parameters $\phi$.
We note that employing a sample from a parametric distribution in (\ref{eq:gof}) does not restrict the ${\cal E}SJS$, and so it is also possible to measure one empirical distribution against any another.

The Bayes factor \cite{KASS95} is a method for model comparison, taking the ratio of the models representing the likelihood of the data under the alternative hypothesis and likelihood of the data under the null hypothesis.
In particular, the Bayes factor is advocated as an alternative method for null hypothesis significance testing, which depends only on the data and considers the models arising from both the null and alternative hypotheses \cite{JARO14}.

The {\em Empirical survival Jensen-Shannon divergence factor} is reformulation of the Bayes factor with the ${\cal E}SJS$, defined as
\begin{equation}\label{eq:jsd-factor}
\frac{{\cal E}SJS(P(D_1,\phi_1), \hat{P})}{{\cal E}SJS(P(D_0,\phi_0), \hat{P})},
\end{equation}
which is the odds ratio of choosing the alternative hypothesis, $P(D_1,\phi_1)$, in preference to the null hypothesis, $P(D_0,\phi_0)$.

\section{Experiments and Analysis}
\label{sec:exp}

To assess the use of the ${\cal E}SJS$ as a goodness-of-fit measure, we provide experimental results with simulated and empirical data with respect to various parametric distributions including the Normal, Uniform, Log-normal, Gamma, Weibull, Beta, Exponential, Pareto \cite{KRIS15} and $q$-Gaussian \cite{NADA07,TSAL17} distributions. It is worth to note that $q$-Gaussian distribution is equivalent to Student's t-distribution \cite{KRIS15}. In the experiments we use the following parametrizations of these distributions:
\begin{itemize}
    \item Normal distribution (two parameters are reported in the Tables: first parameter is the mean $ \mu $ and the second parameter is standard deviation $ \sigma $):
        \begin{equation}
            p_{\mu,\sigma}(x) = {\frac{1}{\sqrt {2\pi \sigma ^{2}}}} \exp\left( -{\frac{(x-\mu )^{2}}{2\sigma ^{2}}} \right) .
        \end{equation}
    \item Uniform distribution (first parameter is the lower bound, $ x_l $, and the second parameter is the upper bound, $ x_u $):
        \begin{equation}
            p_{x_l,x_u}(x) = \left\{ \begin{array}{ll}
                \frac{1}{x_u - x_l} & \text{if } x_l \leq x \leq x_u \\
                0 & \text{otherwise}
            \end{array} \right. .
        \end{equation}
    \item Log-normal distribution (first parameter -- $ \mu $, second parameter -- $ \sigma $):
        \begin{equation}
            p_{\mu,\sigma}(x) = \frac{1}{\sigma \sqrt {2\pi } x} \exp \left(-{\frac{\left(\ln x-\mu \right)^{2}}{2\sigma ^{2}}}\right) .
        \end{equation}
    \item Gamma distribution (first parameter -- $ k $, second parameter -- $ \tau $):
        \begin{equation}
            p_{k,\tau}(x) = \frac{x^{k-1}}{\Gamma (k) \tau^{k}} \exp\left( -\frac{x}{\tau} \right) .
        \end{equation}
    \item Weibull distribution (first parameter -- $ k $, second parameter -- $ \tau $):
        \begin{equation}
            p_{k,\tau}(x) = \frac{k x^{k-1}}{\tau^k} \exp \left(- \frac{x^k}{\tau^k} \right) .
        \end{equation}
    \item Beta distribution (first parameter -- $ \alpha $, second parameter -- $ \beta $):
        \begin{equation}
            p_{\alpha,\beta}(x) = \frac{\Gamma(\alpha) \Gamma(\beta)}{\Gamma(\alpha+\beta)} x^{\alpha-1} (1-x)^{\beta -1} .
        \end{equation}
    \item $q$-Gaussian distribution (first parameter -- $q$, second parameter -- $ \lambda $):
        \begin{equation}
            p_{\lambda,x_0}(x) = \frac{\Gamma\left(\frac{\lambda}{2}\right)}{\sqrt{\pi x_0^2} \Gamma\left(\frac{\lambda-1}{2}\right)} \left(\frac{x_0^2}{x_0^2+x^2}\right)^{\frac{\lambda}{2}} .
        \end{equation}
    \item Exponential distribution (only one parameter is reported -- $ \tau $):
        \begin{equation}
            p_\tau(x) = \frac{1}{\tau} \exp\left( - \frac{x}{\tau} \right) .
        \end{equation}
    \item Pareto distribution (only one parameter is reported -- $ \alpha $):
        \begin{equation}
            p_\alpha(x) = \frac{\alpha}{x^{\alpha+1}} .
        \end{equation}
\end{itemize}

In the experiments we carry out, we will make use of the {\em bootstrap method} \cite{DAVI97}, which is a technique for computing a confidence interval that relies on random resampling with replacement from a given sample data set. The bootstrap method is usually nonparametric, making no distributional assumptions about the data set employed.

Our methodology for the experiments with simulated data (see Subsection~\ref{subsec:simulated}) was as follows:
\renewcommand{\labelenumi}{(\roman{enumi})}
\begin{enumerate}
\item First we generated a data set, say $S_1$, of size $10^6$ from a {\em given distribution}, say $D_1$, with chosen parameters, say $\phi_1$, which was then taken to be the empirical distribution.

\item We then considered $S_1$ to be distributed according to a {\em hypothesised distribution}, $D_2$, where $D_2$ may not be the same as $D_1$, and used the maximum likelihood method to obtain the parameters of $S_1$, say $\phi_2$, assuming its distribution was $D_2$. (Obviously, if $D_2 = D_1$, then $\phi_2$ is expected to be very close to $\phi_1$.)

\item Next, assuming that $S_1$ was distributed according to $D_2$ with parameters $\phi_2$, we generated a second data set, $S_2 = P(D_2, \phi_2)$, from distribution $D_2$ with parameters $\phi_2$.

\item Finally, we evaluated ${\cal E}SJS(S_2, S_1)$ as a measure of the goodness-of-fit of $D_2$, with parameters $\phi_2$, to $S_1$, and computed a 95\% confidence interval for the ${\cal E}SJS$ from 1000 bootstrap resamples using the {\em basic bootstrap percentile method} \cite[Section 5.3.1]{DAVI97}.
\end{enumerate}

For the experiments with empirical data sets (see Subsection~\ref{subsec:empirical}) we followed the same methodology, with the difference that the data set $S_1$ was an empirical data set rather than a generated one. In this case the survival function is estimated via a step function, known as the {\em Kaplan-Meier estimator} \cite{KAPL58,KLEI12}, making use of $10^6$ bins.

For each set of experiments we followed the methodology described above for several possible alternative parametric distributions, $D_2$, and then computed the ${\cal E}SJS$ factor between the best and a lower performing distribution.

\subsection{Experiments with Simulated Data}
\label{subsec:simulated}

We now provide commentary on the results for the simulated data, shown in Tables \ref{table:norm01}, \ref{table:logn01}, \ref{table:gamma22}, \ref{table:gamma502}, \ref{table:beta22}, \ref{table:beta5050} and \ref{table:beta6030}.

In the first experiment the given distribution was Normal with mean $\mu = 0$ and standard deviation $\sigma = 1$, and the hypothesised distributions were Normal and Uniform. The ${\cal E}SJS$ factor between the ${\cal E}SJS$s of the Normal and Uniform distributions is $803.9692$, which can be derived from Table~\ref{table:norm01}.

\begin{table}[!ht]
\begin{center}
\begin{tabular}{|l|c|c|c|c|c|}\hline
Distribution & Parameter 1 & Parameter 2 & ${\cal E}SJS$ & lb     & ub     \\ \hline \hline
Normal       & 0.0003      & 0.9995      & {\bf 0.0002}  & 0.0002 & 0.0005 \\ \hline
Uniform      & -4.7467     & 4.8122      & {\em 0.1947}  & 0.1911 & 0.1950 \\ \hline
\end{tabular}
\end{center}
\caption{\label{table:norm01} Experiment 1: Normal distribution with mean $\mu = 0$ and standard deviation $\sigma = 1$.}
\end{table}

In the second experiment the given distribution was Log-normal with mean $\mu = 0$ and standard deviation $\sigma = 1$, and the hypothesised distributions were Normal, Uniform, Log-normal, Gamma and Weibull. The ${\cal E}SJS$ factor between the ${\cal E}SJS$s of the Log-normal distribution, which is the smallest, and the Gamma distribution, whose ${\cal E}SJS$ is the closest to it, is $207.5271$, which can be derived from Table~\ref{table:logn01}.

\begin{table}[!ht]
\begin{center}
\begin{tabular}{|l|c|c|c|c|c|}\hline
Distribution & Parameter 1 & Parameter 2 & ${\cal E}SJS$ & lb     & ub     \\ \hline \hline
Normal       & 1.6492      & 2.1497      & 0.1520        & 0.1513 & 0.1527 \\ \hline
Uniform      & 0.0054      & 104.6298    & 0.9001        & 0.8816 & 0.9005\\ \hline
Log-normal   & 0.0001      & 1.0006      & {\bf 0.0002}  & 0.0002 & 0.0005 \\ \hline
Gamma        & 1.1373      & 1.4501      & {\em 0.0481}  & 0.0477 & 0.0484 \\ \hline
Weibull      & 1.0002      &  1.6494     & 0.0543        & 0.0540 & 0.0547 \\ \hline
\end{tabular}
\end{center}
\caption{\label{table:logn01} Experiment 2: Log-normal distribution with mean $\mu = 0$ and standard deviation $\sigma = 1$.}
\end{table}

In the third experiment the given distribution was Gamma with shape $\alpha = 2$ and scale $\theta = 2$, and the hypothesised distributions were Normal, Uniform, Log-normal, Weibull and Gamma. The ${\cal E}SJS$ factor between the ${\cal E}SJS$s of the Gamma distribution, which is the smallest, and the Weibull distribution whose ${\cal E}SJS$ is the closest to it, is $85.8914$, which can be derived from Table~\ref{table:gamma22}.

\begin{table}[!ht]
\begin{center}
\begin{tabular}{|l|c|c|c|c|c|}\hline
Distribution & Parameter 1 & Parameter 2 & ${\cal E}SJS$ & lb     & ub     \\ \hline \hline
Normal       & 3.9999      & 2.8282      & 0.0743        & 0.0741 & 0.0746 \\ \hline
Uniform      & 0.0020      & 34.4431     & 0.5539        & 0.5137 & 0.5543 \\ \hline
Log-normal   & 1.1164      & 0.8020      & 0.0308        & 0.0305 & 0.0311 \\ \hline
Gamma        & 2.0031      & 1.9969      & {\bf 0.0002}  & 0.0002 & 0.0005 \\ \hline
Weibull      & 1.4831      & 4.4386      & {\em 0.0150}  & 0.0147 & 0.0153 \\ \hline
\end{tabular}
\end{center}
\caption{\label{table:gamma22} Experiment 3: Gamma distribution with shape $\alpha = 2$ and scale $\theta = 2$.}
\end{table}

In the fourth experiment the given distribution was Gamma with shape $\alpha = 50$ and scale $\theta = 2$, and the hypothesised distributions were Normal, Uniform, Log-normal, Weibull and Gamma. The ${\cal E}SJS$ factor between the ${\cal E}SJS$s of the Gamma distribution, which is the smallest, and the Log-normal distribution whose ${\cal E}SJS$ is the closest to it, is $17.9510$, which can be derived from Table~\ref{table:gamma502}.

\begin{table}[!ht]
\begin{center}
\begin{tabular}{|l|c|c|c|c|c|}\hline
Distribution & Parameter 1 & Parameter 2 & ${\cal E}SJS$ & lb     & ub     \\ \hline \hline
Normal       & 100.0013    & 14.1418     & 0.0132        & 0.0130 & 0.0135  \\ \hline
Uniform      & 44.9060     & 186.2970    & 0.2073        & 0.2019 & 0.2103  \\ \hline
Log-normal   & 4.5952      & 0.1421      & {\em 0.0059}  & 0.0057 & 0.0062  \\ \hline
Gamma        & 50.0161     & 1.9994      & {\bf 0.0003}  & 0.0002 & 0.0006  \\ \hline
Weibull      &  7.2817     & 106.2044    & 0.04636       & 0.0460 & 0.04667 \\ \hline
\end{tabular}
\end{center}
\caption{\label{table:gamma502} Experiment 4: Gamma distribution with shape $\alpha = 50$ and scale $\theta = 2$.}
\end{table}

In the fifth experiment the given distribution was Beta with parameters $\alpha = 2$ and $\beta = 2$, and the hypothesised distributions were Normal, Log-normal, Gamma, Weibull and Beta. The ${\cal E}SJS$ factor between the Beta distribution which is the smallest, and the Normal distribution, whose ${\cal E}SJS$ is the closest to it, is $99.0703$, which can be derived from Table~\ref{table:beta22}.

\begin{table}[!ht]
\begin{center}
\begin{tabular}{|l|c|c|c|c|c|}\hline
Distribution & Parameter 1 & Parameter 2 & ${\cal E}SJS$  & lb     & ub     \\ \hline \hline
Normal       & 0.4999      & 0.2237      & {\em 0.0246}   & 0.0244 & 0.0248 \\ \hline
Log-normal   & -0.8340     & 0.6020      & 0.0583         & 0.0581 & 0.0586 \\ \hline
Gamma        & 3.7158      & 0.1345      & 0.0399         & 0.0396 & 0.0401 \\ \hline
Beta         & 1.9973      & 1.9988      & {\bf 0.0002}   & 0.0002 & 0.0005 \\ \hline
Weibull      & 2.3816      & 0.5628      & 0.0255         & 0.0253 & 0.0258 \\ \hline
\end{tabular}
\end{center}
\caption{\label{table:beta22} Experiment 5: Beta distribution with $\alpha = 2$ and $\beta = 2$.}
\end{table}

In the sixth experiment the given distribution was Beta with parameters $\alpha = 50$ and $\beta = 50$, and the hypothesised distributions were Normal, Log-normal, Gamma, Weibull and Beta. The ${\cal E}SJS$ factor between the Beta distribution, which is the smallest, and the Normal distribution, whose ${\cal E}SJS$ is the closest to it, is $6.4432$, which can be derived from Table~\ref{table:beta5050}.

\begin{table}[!ht]
\begin{center}
\begin{tabular}{|l|c|c|c|c|c|}\hline
Distribution & Parameter 1 & Parameter 2 & ${\cal E}SJS$ & lb     & ub    \\ \hline \hline
Normal       & 0.5000      & 0.0497      & {\em 0.0010}  & 0.0008 & 0.0013 \\ \hline
Log-normal   & -0.6982     & 0.1007      & 0.0128        & 0.0125 & 0.0131 \\ \hline
Gamma        & 99.7537     & 0.0050      & 0.0086        & 0.0083 & 0.0088 \\ \hline
Beta         & 50.0429     & 50.0496     & {\bf 0.0002}  & 0.0002 & 0.0004 \\ \hline
Weibull      & 10.6949     & 0.5225      & 0.0369        & 0.0365 & 0.0372 \\ \hline
\end{tabular}
\end{center}
\caption{\label{table:beta5050} Experiment 6: Beta distribution with $\alpha = 50$ and $\beta = 50$.}
\end{table}

In the seventh and final experiment the given distribution was Beta with parameters $\alpha = 60$ and $\beta = 30$, and the hypothesised distributions were Normal, Log-normal, Gamma, Weibull  and Beta. The ${\cal E}SJS$ factor between the Beta distribution, which is the smallest, and the Normal distribution, whose ${\cal E}SJS$ is the closest to is, is $30.8739$, which can be derived from Table~\ref{table:beta6030}.

\begin{table}[!ht]
\begin{center}
\begin{tabular}{|l|c|c|c|c|c|}\hline
Distribution & Parameter 1 & Parameter 2 & ${\cal E}SJS$ & lb     & ub     \\ \hline \hline
Normal       & 0.6666      & 0.0494      & {\em 0.0064}  & 0.0062 & 0.0067 \\ \hline
Log-normal   & -0.4084     & 0.0750      & 0.0158        & 0.0156 & 0.0161 \\ \hline
Gamma        & 179.5350    & 0.0037      & 0.0127        & 0.0125 & 0.0130 \\ \hline
Beta         & 60.0885     & 30.0540     & {\bf 0.0002}  & 0.0002 & 0.0005 \\ \hline
Weibull      & 14.6275     & 0.6893      & 0.0326        & 0.0323 & 0.0329 \\ \hline
\end{tabular}
\end{center}
\caption{\label{table:beta6030} Experiment 7: Beta distribution with $\alpha = 60$ and $\beta = 30$.}
\end{table}

We note that all the computations described in this subsection were carried out using the Matlab software package. Table~\ref{table:jsd-factors} summarises, for all experiments, the ${\cal E}SJS$ factors between the best (highlighted in bold) and second best (highlighted in italics) performing distributions. In all cases, apart from experiment 6 for the Beta distribution with $\alpha = 50$ and $\beta = 50$, shown in Table~\ref{table:beta5050}, the ${\cal E}SJS$ factor overwhelmingly supports the given distribution, as we would expect. The reason for the relatively low ${\cal E}SJS$ factor in this particular case is the known fact that the Beta distribution can be approximated by the Normal distribution when $\alpha$ and $\beta$ are large \cite{PEIZ68}.

\begin{table}[!ht]
\begin{center}
\begin{tabular}{|l|c|c|}\hline
Experiment     & Distribution & ${\cal E}SJS$ factor \\ \hline \hline
1              & Normal       & 803.9692 \\ \hline
2              & Log-normal   & 270.5271 \\ \hline
3              & Gamma        & 85.8914  \\ \hline
4              & Gamma        & 17.9510  \\ \hline
5              & Beta         & 99.0703  \\ \hline
6              & Beta         & 6.4432   \\ \hline
7              & Beta         & 30.8739  \\ \hline
\end{tabular}
\end{center}
\caption{\label{table:jsd-factors} The ${\cal E}SJS$ factors between the best and second best performing distributions for the simulated data.}
\end{table}

\begin{figure}[!ht]
\begin{center}
\includegraphics[width=0.9\textwidth]{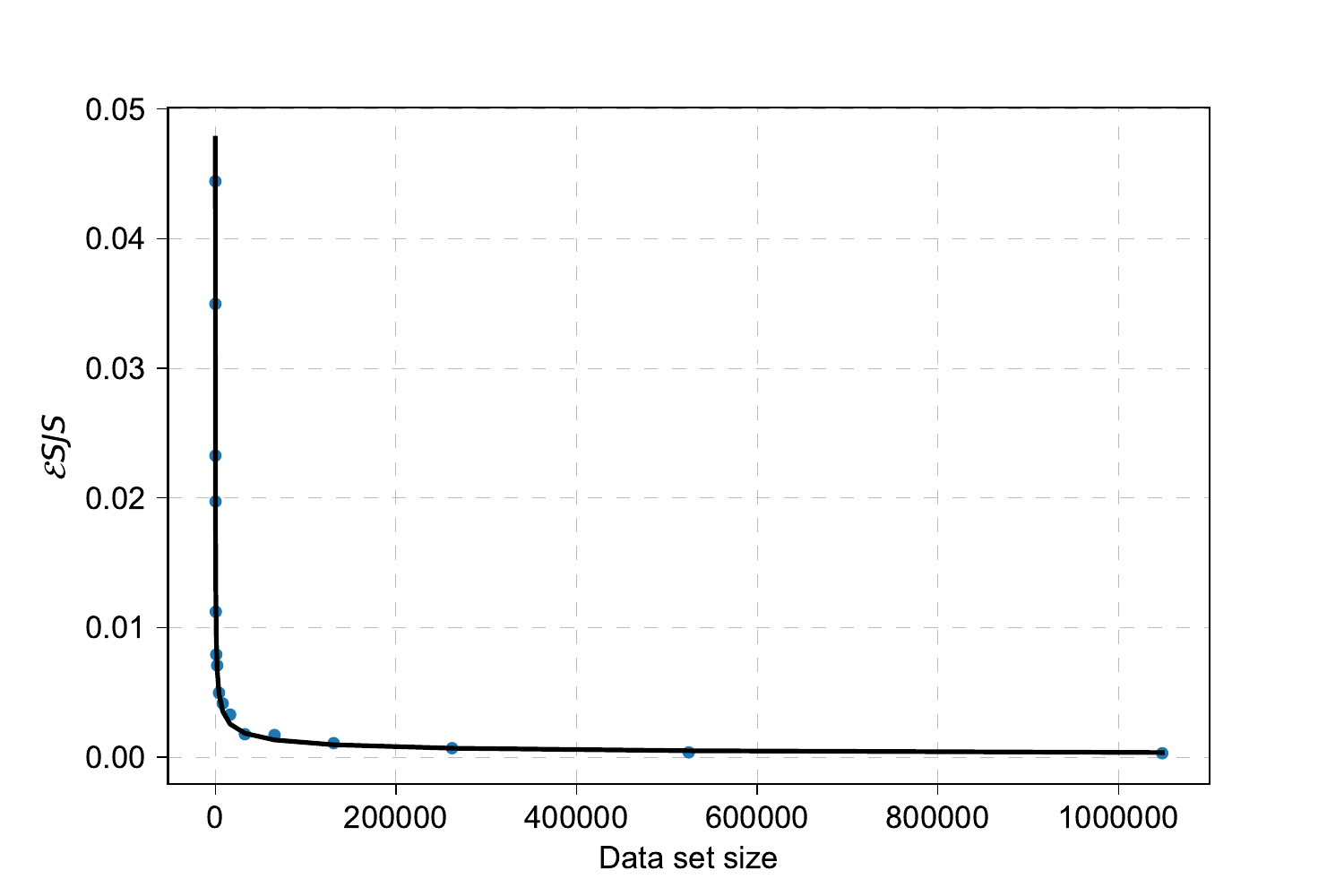}
\caption{\label{figure:pl} Power-law fit, $0.2426 x^{-0.4691}$, for data in Table~\ref{table:norm01-sz}.}
\end{center}
\end{figure}

It is evident that the larger the size of the data set the more accurate the ${\cal E}SJS$ will be.
In Table~\ref{table:norm01-sz} we demonstrate how the accuracy of the ${\cal E}SJS$ increases while the data set size increases,
when both the given and hypothesised distribution are Normal with mean $\mu = 0$ and standard deviation $\sigma = 1$; it is also noticeable that the maximum likelihood estimation of parameters converges to the correct one as the size of the data increases.
In Figure~\ref{figure:pl} we show that the decrease of the ${\cal E}SJS$, empirically, follows a power-law distribution with an exponent of approximately 0.5, which is $\sqrt{x}$, where $x$ represents the data size.

\begin{table}[!ht]
\begin{center}
\begin{tabular}{|c|c|c|c|} \hline
Parameter 1 & Parameter 2 & $ {\cal E}SJS $ & Data set size \\ \hline \hline
-0.3125    & 0.9571        & 0.0444          & 32            \\ \hline
-0.0171    & 1.0037        & 0.0350          & 64            \\ \hline
 0.0098    & 1.0424        & 0.0233          & 128           \\ \hline
-0.0153    & 1.0361        & 0.0197          & 256           \\ \hline
-0.0119    & 1.0407        & 0.0112          & 512           \\ \hline
-0.0022    & 1.0049        & 0.0079          & 1024          \\ \hline
-0.0140    & 0.9894        & 0.0071          & 2048          \\ \hline
-0.0084    & 0.9915        & 0.0050          & 4096          \\ \hline
 0.0178    & 0.9819        & 0.0042          & 8192          \\ \hline
 0.0091    & 0.9895        & 0.0033          & 16384         \\ \hline
 0.0090    & 0.9924        & 0.0018          & 32768         \\ \hline
 0.0076    & 0.9950        & 0.0017          & 65536         \\ \hline
 0.0053    & 0.9936        & 0.0011          & 131072        \\ \hline
 0.0061    & 0.9962        & 0.0007          & 262144        \\ \hline
 0.0028    & 0.9976        & 0.0004          & 524288        \\ \hline
 0.0012    & 0.9992        & 0.0003          & 1048576       \\ \hline
\end{tabular}
\end{center}
\caption{\label{table:norm01-sz}  The ${\cal E}SJS$ resulting from increasing the data set size, where the given distribution is Normal with mean $\mu = 0$ and standard deviation $\sigma = 1$.}
\end{table}

\subsection{Analysis of Empirical Data}
\label{subsec:empirical}

We now provide commentary on the results for the empirical data sets, shown in Tables \ref{table:LR1992}, \ref{table:financialData} and  \ref{table:football}.
We note that all the computations described in this subsection were carried out using Python.
Table~\ref{table:emp-jsd-factors} summarises, for all three data sets, the ${\cal E}SJS$ factors between the best (highlighted in bold) and a lower performing distribution (highlighted in italics).

The first empirical data set we consider, contains detailed voting results of party
vote shares in different polling stations, during the Lithuanian parliamentary
election of 1992 (the data was obtained from \cite{KONO16});
for each party under consideration we have examined $2061$ data points.
For this data set we use the interval $(0,1)$ for quantifying the vote shares,
since these are naturally bounded to this value range.

Note that we consider only the top three parties and have renormalised the original
data so that the total vote share of the top three parties would sum to one in each polling station.
This data set was first considered in \cite{KONO17}, where an agent-based model generating the Beta distribution, and reasonably
well reproducing detailed election results, was proposed. In \cite{KONO18} a statistical comparison between the four commonly used distributions in
sociophysics \cite{SEN14}, the Normal, Log-normal, Beta and Weibull, was carried out using the Watanabe-Akaike information criterion ({\em WAIC}) \cite{WATA13}, which is a generalisation of the AIC.
The comparison concluded that the Beta and Weibull distributions provide the best fits for the empirical data.
However, their respective WAIC scores were within each other's confidence intervals, and therefore no final conclusion was made.
Here we also obtain a similar result, the Beta and Weibull distributions clearly have the overall best scores, however, as before, their confidence intervals overlap (see Table~\ref{table:LR1992}).
As was noted in \cite{KONO18}, the Beta and Weibull distributions are similar when the observed mean is close to $0.5$
and the observed variance is reasonably small. In the empirical analysis this similarity is further increased when the sample size is small.
In addition, for the estimated parameter values, the Gamma and Weibull distributions behave similarly when $x \in (0,1)$.
Therefore, we report the ${\cal E}SJS$ factor between the best performing distribution (highlighted in bold) and the
next best distribution which is neither a Beta, Gamma nor a Weibull distribution (see Table~\ref{table:emp-jsd-factors}).

\begin{table}[!ht]
\begin{center}
\begin{tabular}{| l | c | c | c | c | c |} \hline
Distribution & Parameter 1 & Parameter 2 & ${\cal E}SJS$ & lb & ub \\ \hline \hline
\multicolumn{6}{|c|}{\bf {SK -- S\k{a}j\={u}d\v{z}io koalicija}} \\ \hline
Normal     & 0.2412  & 0.1132 & 0.0265       & 0.0229 & 0.0307 \\ \hline
Log-normal & -1.5556 & 0.5721 & {\em 0.0243} & 0.0203 & 0.0290 \\ \hline
Gamma      & 3.9083  & 0.0617 & 0.0126       & 0.0097 & 0.0161 \\ \hline
Beta       & 3.0771  & 9.7045 & 0.0091       & 0.0067 & 0.0125 \\ \hline
Weibull    & 2.2492  & 0.2722 & {\bf 0.0089} & 0.0068 & 0.0127 \\ \hline \hline
\multicolumn{6}{|c|}{{\bf LKDP -- Lietuvos krik\v{s}\v{c}ioni\k{u} demokrat\k{u} partija}} \\ \hline
Normal     & 0.1575  & 0.0921  & 0.0308       & 0.0276 & 0.0339 \\ \hline
Log-normal & -2.0435 & 0.6884  & {\em 0.0168} & 0.0138 & 0.0202 \\ \hline
Gamma      & 2.7196  & 0.0579  & 0.0060       & 0.0044 & 0.0084 \\ \hline
Beta       & 2.3282  & 12.4475 & {\bf 0.0057} & 0.0044 & 0.0082 \\ \hline
Weibull    & 1.7869  & 0.1773  & 0.0081       & 0.0060 & 0.0110 \\ \hline \hline
\multicolumn{6}{|c|}{\bf{LDDP -- Lietuvos demokratin\.{e} darbo partija}} \\ \hline
Normal     & 0.6013  & 0.1516 & {\em 0.0169} & 0.0135 & 0.0236 \\ \hline
Log-normal & -0.5459 & 0.2892 & 0.0525       & 0.0468 & 0.0582 \\ \hline
Gamma      & 13.5242 & 0.0445 & 0.0423       & 0.0361 & 0.0485 \\ \hline
Beta       & 5.4454  & 3.6014 & 0.0106       & 0.0093 & 0.0181 \\ \hline
Weibull    & 4.5344  & 0.6588 & {\bf 0.0089} & 0.0081 & 0.0172 \\ \hline
\end{tabular}
\end{center}
\caption{\label{table:LR1992} Empirical data set 1: Lithuanian parliamentary election 1992.}
\end{table}

The second data set we consider contains the log-returns of two different exchange rates.
We consider the BTC/JPY exchange rate during the time period between July 4, 2017
and July 4, 2018  on the bitFlyer exchange (the data was obtained from \cite{BTCData}),
as well as the EUR/USD exchange rate during the time period between June 1, 2000 and September 1, 2010
(the data was obtained from \cite{ForexData}).
We have taken into consideration the daily and one minute log-returns.
For this data set we use the interval $[-100,100]$ to quantify the returns for the unbounded distributions (Normal and $q$-Gaussian distributions),
and the interval $[0.01,100]$ to quantify the returns for the bounded distributions (Log-normal, Gamma and Weibull distributions).
We have selected this value range, since the maximum observed absolute return value
in this data set is approximately $85$. The number of return data points we examine is as follows:
$364$ for the daily BTC/JPY exchange rate, $526541$ for the 1 min BTC/JPY exchange rate,
$2544$ for the daily EUR/USD exchange rate and $3665755$ for the 1 min EUR/USD exchange rate.
For this data set we employ the moving block bootstrap \cite{KREI12} with a block size of one day.

In the econophysics literature \cite{SLAN14} it is commonly accepted that the log-returns are power-law distributed \cite{CONT01}.
One of the commonly used fits for the log-returns is so-called
$q$-Gaussian distribution \cite{TSAL17}, which we add to
our analysis for this empirical data set.

However, as can be seen in Table~\ref{table:financialData}, we find that
the Gamma and Weibull distributions noticeably outperform the
$q$-Gaussian distribution. Performance of the Gamma and Weibull
distributions is similar, due to the fact that for the estimated parameter values
both of these distributions behave reasonably similarly;
for these parameter values they are reasonably close to the Exponential distribution.
Therefore, we report the ${\cal E}SJS$ factor between the best performing distribution (highlighted in bold) and the
next best distribution, which is neither a Gamma nor a Weibull distribution  (see Table~\ref{table:emp-jsd-factors}).

For our fourth sample in the second empirical data set, i.e. the EUR/USD one minute log-returns, unexpectedly the Log-normal
and $q$-Gaussian distributions had the best performance. Though they are far from being
similar for the considered observable value range and parameter values, they, most likely
attained similar scores due to the shape of the empirical distribution.
The Log-normal distribution seems to represent smaller log-returns well,
while the $q$-Gaussian is better at describing the tail events.

\begin{table}[!ht]
\begin{center}
\begin{tabular}{| l | c | c | c | c | c |} \hline
Distribution & Parameter 1 & Parameter 2 & ${\cal E}SJS$ & lb & ub \\ \hline \hline
\multicolumn{6}{|c|}{{\bf Daily -- BTC/JPY (bitFlyer)}} \\ \hline
Normal     & 0.0415  & 1.0000 & 0.0099       & 0.0055 & 0.0144 \\ \hline
Log-normal & -0.9039 & 1.1431 & {\em 0.0030} & 0.0017 & 0.0050 \\ \hline
Gamma      & 1.1015  & 0.6174 & {\bf 0.0020} & 0.0010 & 0.0028 \\ \hline
Weibull    & 1.0276  & 0.6882 & 0.0024       & 0.0010 & 0.0030 \\ \hline
$q$-Gaussian & 3.9028  & 1.0413 & 0.0061       & 0.0048 & 0.0116 \\ \hline \hline
\multicolumn{6}{|c|}{{\bf One minute -- BTC/JPY (bitFlyer)}} \\ \hline
Normal     & 0.0011  & 1.0000 & 0.0140       & 0.0121 & 0.0158 \\ \hline
Log-normal & -1.2561 & 1.5771 & {\em 0.0075} & 0.0070 & 0.0077 \\ \hline
Gamma      & 0.7717  & 0.7993 & {\bf 0.0033} & 0.0026 & 0.0029 \\ \hline
Weibull    & 0.8468  & 0.5654 & 0.0034       & 0.0028 & 0.0031 \\ \hline
$q$-Gaussian & 2.8964  & 0.5871 & 0.0082       & 0.0069 & 0.0097 \\ \hline \hline
\multicolumn{6}{|c|}{{\bf Daily -- EUR/USD (Forex)}} \\ \hline
Normal     & 0.0179  & 1.0000 & 0.0038       & 0.0026 & 0.0053 \\ \hline
Log-normal & -0.7227 & 1.1086 & 0.0049       & 0.0042 & 0.0052 \\ \hline
Gamma      & 0.6102  & 1.2489 & 0.0010       & 0.0007 & 0.0014 \\ \hline
Weibull    & 1.1579  & 0.8021 & {\bf 0.0006} & 0.0004 & 0.0011 \\ \hline
$q$-Gaussian & 7.8686  & 2.2146 & {\em 0.0022} & 0.0018 & 0.0039 \\ \hline
\hline
\multicolumn{6}{|c|}{{\bf One minute -- EUR/USD (Forex)}} \\ \hline
Normal     & 0.0004  & 1.0000 & 0.0138       & 0.0136 & 0.0142 \\ \hline
Log-normal & -0.3236 & 0.6165 & {\bf 0.0086} & 0.0085 & 0.0087 \\ \hline
Gamma      & 2.3423  & 0.3881 & 0.0116       & 0.0113 & 0.0116 \\ \hline
Weibull    & 1.3548  & 1.0057 & 0.0145       & 0.0142 & 0.0144 \\ \hline
$q$-Gaussian & 2.9820  & 0.6415 & {\em 0.0092} & 0.0087 & 0.0093 \\ \hline
\end{tabular}
\end{center}
\caption{\label{table:financialData} Empirical Data Set 2: Daily and one minute log-returns of EUR/USD and BTC/JPY exchange rates.}
\end{table}

Similar conclusions follow from the WAIC scores. Using the same methodology as in \cite{KONO18}
we have also calculated WAIC scores for the second empirical data set. For the first three samples
in the data set, we observe that the Gamma and Weibull distributions have the best WAIC scores and, in addition, have overlapping confidence intervals. Moreover, in all three cases, the Log-normal distribution has the next best WAIC score. For the fourth sample the Log-normal distribution
has the lowest score, while the $q$-Gaussian distribution is the next best.

The third data set we consider is the European soccer data set \cite{EuroFootData},
which contains $25$ thousand matches played in European national championships throughout 2008--2016.
From this data set we have extracted five random teams and computed inter--goal times for each team.
We have used the interval $[1,505]$ to quantify the inter--gaol times, since the longest inter--goal time observed in the considered series was $501$ minutes.
The number of inter--goal times under consideration for each team is as follows:
$471$  for TOT, $397$  for GLA, $564$  for MUN, $475$  for VAL and $64$ for ELC.

We introduce two new distributions for the third data set analysis. First of all
we include exponential distribution, which should arise if goal scoring rate is
independent of previous goal scoring history. Yet, if goal scoring rate is increasing
as more goals are scored, we could expect to observed power--law distribution.
Power--law inter--event time distributions are observed in many social systems, e.g.,
human communication time series exhibit such pattern \cite{BARA05}.

\begin{table}[!htp]
\begin{center}
\begin{tabular}{| l | c | c | c | c | c |}  \hline
Distribution & Parameter 1 & Parameter 2 & ${\cal E}SJS$ & lb & ub \\ \hline \hline
\multicolumn{6}{|c|}{{\bf TOT -- Tottenham Hotspur (English Premier League)}} \\ \hline
Normal      & 57.7346 & 58.2559 & 0.0381       & 0.0326 & 0.0423 \\ \hline
Gamma       & 1.0565  & 54.6477 & 0.0071       & 0.0055 & 0.0109 \\ \hline
Weibull     & 1.0183  & 58.1866 & 0.0069       & 0.0054 & 0.0102 \\ \hline
Exponential & 57.7346 & -       & {\bf 0.0066} & 0.0057 & 0.0115 \\ \hline
Pareto      & 1.6809  & -       & {\em 0.0249} & 0.0200 & 0.0310 \\ \hline \hline
\multicolumn{6}{|c|}{{\bf GLA -- Borussia Monchengladbach (German Bundesliga)}} \\ \hline
Normal      & 61.5088 & 61.3039 & 0.0390       & 0.0334 & 0.0424 \\ \hline
Gamma       & 1.0186  & 60.3877 & 0.0075       & 0.0063 & 0.0111 \\ \hline
Weibull     & 1.0012  & 61.5400 & 0.0072       & 0.0062 & 0.0106 \\ \hline
Exponential & 61.5088 & -       & {\bf 0.0071} & 0.0062 & 0.0127 \\ \hline
Pareto      & 1.6850  & -       & {\em 0.0235} & 0.0187 & 0.0303 \\ \hline \hline
\multicolumn{6}{|c|}{{\bf MUN -- Manchester United (English Premier League)}} \\ \hline
Normal      & 48.3174 & 49.8385 & 0.0366       & 0.0315 & 0.0394 \\ \hline
Gamma       & 1.0497  & 46.0306 & 0.0070       & 0.0051 & 0.0101 \\ \hline
Weibull     & 1.0084  & 48.4960 & 0.0068       & 0.0050 & 0.0094 \\ \hline
Exponential & 48.3174 & -       & {\bf 0.0067} & 0.0050 & 0.0103 \\ \hline
Pareto      & 1.6886  & -       & {\em 0.0242} & 0.0203 & 0.0301 \\ \hline
\hline
\multicolumn{6}{|c|}{{\bf VAL -- Valencia CF (Spanish La Liga)}} \\ \hline
Normal      & 57.2779 & 52.8406 & 0.0334       & 0.0284 & 0.0365 \\ \hline
Gamma       & 1.1212  & 51.0879 & {\bf 0.0053} & 0.0048 & 0.0090 \\ \hline
Weibull     & 1.0725  & 58.8645 & 0.0054       & 0.0047 & 0.0093 \\ \hline
Exponential & 57.2779 & -       & 0.0066       & 0.0051 & 0.0122 \\ \hline
Pareto      & 1.6547  & -       & {\em 0.0287} & 0.0238 & 0.0347 \\ \hline \hline
\multicolumn{6}{|c|}{{\bf ELC -- Elche CF (Spanish La Liga/La Liga 2)}} \\ \hline
Normal      & 102.1875 & 87.1523  & 0.0402       & 0.0263 & 0.0658 \\ \hline
Gamma       & 1.3220   & 77.2953  & 0.0173       & 0.0161 & 0.0349 \\ \hline
Weibull     & 1.1929   & 108.4610 & {\bf 0.0158} & 0.0157 & 0.0332 \\ \hline
Exponential & 102.1875 & -        & 0.0281       & 0.0192 & 0.0581 \\ \hline
Pareto      & 1.6154   & -        & {\em 0.0400} & 0.0257 & 0.0632 \\ \hline
\end{tabular}
\end{center}
\caption{\label{table:football} Empirical data set 3: Inter--goal times recorded in the European soccer data set.}
\end{table}

We have treated goals scored during extra time as scored on the 45th minute (if scored during the first half) and the 90th minute (if scored
during the second half). In this analysis we have added the Exponential and Pareto distributions.
For the estimated parameter values the Gamma and Weibull distributions behave similarly to the Exponential distribution.
Note that the shape parameter values of the Gamma and Weibull distributions are very
close to $1$ and the respective scale parameter values are similar.
In this case it is known that Gamma and Weibull distributions are equivalent to the Exponential distribution with the appropriate scale parameter value.
We therefore report the ${\cal E}SJS$ factor between the best performing distribution (highlighted in bold) and the
next best distribution, which is neither an Exponential, Gamma nor a Weibull distribution  (see Table~\ref{table:emp-jsd-factors}).
We observe that for the ELC sample, the obtained ${\cal E}SJS$ factor is the lowest and the ${\cal E}SJS$ score is the largest.
This is most likely due to this team having played opponents with a larger variety of skill. In particular, it played in the top and the second tiers of the national championship during the considered time period, resulting in a goal scoring rate with a higher variation.

Similar conclusions follow from the WAIC scores. Using the same methodology as in \cite{KONO18}
we have also calculated WAIC scores for the third empirical data set. For all samples in the third the data set, 
we observe that the Gamma, Weibull and Exponential distributions have the best WAIC scores and, in addition, have overlapping confidence intervals.
We note that, for the ELC sample, using WAIC, we have not observed any significant differences between the distributions; this is likely, since the sample size for the ELC is small (i.e. $64$ goals).

\begin{table}[!htp]
\begin{center}
\begin{tabular}{| l | c | c | c |} \hline
Data set & Sample         & Distribution & ${\cal E}SJS$ factor \\ \hline \hline
1        & SK             & Weibull      & 2.7157 \\ \hline
1        & LKDP           & Beta         & 2.9374 \\ \hline
1        & LDDP           & Weibull      & 1.8943 \\ \hline \hline
2        & Daily BTC/JPY  & Gamma        & 1.5027 \\ \hline
2        & 1 min  BTC/JPY & Gamma        & 2.2476 \\ \hline
2        & Daily EUR/USD  & Weibull      & 3.3437 \\ \hline
2        & 1 min EUR/USD  & Log-normal   & 1.0564 \\ \hline \hline
3        & TOT            & Exponential  & 3.7628 \\ \hline
3        & GLA            & Exponential  & 3.2758 \\ \hline
3        & MUN            & Exponential  & 3.6089 \\ \hline
3        & VAL            & Weibull      & 5.2841 \\ \hline
3        & ELC            & Weibull      & 2.5242 \\ \hline
\end{tabular}
\end{center}
\caption{\label{table:emp-jsd-factors} The ${\cal E}SJS$ Factors between the best and a lower performing distribution for the empirical data sets.}
\end{table}

\section{Concluding remarks}
\label{sec:conc}

We have proposed the empirical survival Jensen-Shannon divergence (${\cal E}SJS$) as a goodness-of-fit measure for data fitted with maximum likelihood estimation or curve fitting. Our experiments with simulated and empirical data in Section~\ref{sec:exp}, for a variety of parametric distributions
commonly employed in sociophysics and econophysics, show that for simulated data the method is unequivocal in its preference for the true distribution (see Subsection~\ref{subsec:simulated}), and for empirical data the method is effective in selecting the more likely distributions from a selection of hypothesised distributions (see Subsection~\ref{subsec:empirical}).

As we have shown in Section~\ref{sec:jsd} the ${\cal E}SJS$ can be formally defined, building on the survival entropy as a generalisation of the standard Jensen-Shannon divergence, and the ${\cal E}SJS$ factor has an intuitive meaning in terms of an odds ratio, in analogy to the Bayes factor.
Moreover, the implementation of the ${\cal E}SJS$ as a measure of goodness-of-fit or for model comparison is relatively straightforward; see \cite{KONO18a} for a Python implementation of the ${\cal E}SJS$.

Ultimately, for a definitive assessment of how the ${\cal E}SJS$ performs in practice, an extensive comparison with existing goodness-of-fit measures should be carried out, and more experience with empirical data sets is needed.


\end{document}